\documentclass[preprint,aps,amssymb,showpacs]{revtex4}
\begin{document}
\preprint{IMSc/2003/05/10}

\title{The Landau electron problem on a cylinder}

\author{G. Date}
\email{shyam@imsc.res.in}
\affiliation{The Institute of Mathematical Sciences\\
CIT Campus, Chennai-600 113, INDIA.}

\author{P. P. Divakaran}
\email{ppd@imsc.res.in}
\affiliation{The Institute of Mathematical Sciences\\
CIT Campus, Chennai-600 113, INDIA.}
\affiliation{Chennai Mathematical Institute\\
92, G. N. Chetty Road, T. Nagar, Chennai-600 017, INDIA.}
\begin{abstract}
We consider the quantum mechanics of an electron confined to move on an
infinite cylinder in the presence of a uniform radial magnetic field. This
problem is in certain ways very similar to the corresponding problem on
the infinite plane. Unlike the plane however, the group of symmetries
of the magnetic field, namely, rotations about the axis and the axial
translations, is {\em not} realized by the quantum electron but only 
a subgroup comprising rotations and discrete translations along the axial 
direction, is. The basic step size of discrete translations is
such that the flux through the `unit cylinder cell' is quantized in
units of the flux quantum. The result is derived in two different ways:
using the condition of projective realization of symmetry groups and
using the more familiar approach of determining the symmetries of a
given Hamiltonian.
\end{abstract}

\pacs{02.20, 03.65.-w}

\maketitle

\section{Introduction}

The classic `Landau electron' problem deals with the quantum mechanics
of an electron on an infinite plane with a uniform magnetic field
perpendicular to the plane. It has been thoroughly studied for decades and
is the basis for many experimental investigations notably in the
context of the Hall effect. Apart from its practical relevance, it also
exhibits many theoretically interesting features. It is exactly solvable
with a simple spectrum for the Hamiltonian; it has infinite degeneracy
for all energy eigenvalues; this degeneracy is organized in an irreducible 
representation of a Heisenberg group and so on. Interestingly the Heisenberg
group arises as the central extension of the translation subgroup of the 
symmetry group of the plane by the $U(1)$ group of phases, the central 
extension being determined by the magnetic field. This is also a dramatic 
illustration of the Wigner theorem \cite{wigner} namely, that symmetries are 
realized projectively in quantum mechanics i.e. the quantum state space 
carries a projective unitary representation (PUR) of a symmetry group. It is 
dramatic in the sense that a free particle on the plane realizes the translation
symmetry by a unitary representation (UR) but if it is charged, the
presence of a uniform magnetic field forces a non-trivial projective 
realization.

In this work we consider a variant of the Landau electron problem 
by replacing the infinite plane by an infinite cylinder. The magnetic field
is again uniform and perpendicular to the surface of the cylinder, i.e. is 
radially directed and constant in magnitude. This is again exactly
solvable and displays the classic features of the planar case but with
an interesting twist. {\em The invariance of the magnetic field under 
continuous translations along the axial direction cannot be realized
by the quantum electron. However, a discrete subgroup is realized as a
symmetry.} The basic reason for this is that there are simply {\em no} 
non-trivial, continuous, projective, unitary representations of the cylinder 
group, $SO(2) \times \mathbb{R}$. Its subgroup $ SO(2) \times \mathbb{Z}$ 
however does have such (infinite dimensional) representations with properties 
analogous to the planar case. The step size of the discrete translation is 
controlled by the magnetic field so as to make the flux through the
`unit cylinder cell' an integral multiple of the flux quantum.

The surprising nature of the result prompted us to reanalyze the problem 
using the usual methods based on a Hamiltonian involving the gauge
potential explicitly and looking for its symmetries. The same result,
including the condition on the step size, is obtained again. The mechanism 
is a subtle interplay of gauge invariance and the definition of symmetry in 
the context of external gauge potentials. The non-contractible nature of the 
cylinder implies that there is a 1-parameter family of gauge-equivalence 
classes, all giving the same magnetic field. Within each class, {\em only} the 
discrete subgroup of the cylinder group acts as a symmetry group. 

We would like to clarify at the outset our approach. It is well known
that when the configuration space of a system is topologically
non-trivial (non-contractible), there is {\em no unique} quantization
procedure for that classical system \cite{GenQuant}. Many of these are
usually in the canonical approach based on the classical phase space. In
this paper however our approach is somewhat different. 

We identify a group of symmetries of the configuration space which we
expect to be realized in the state space whatever be the quantization 
procedure adopted. The symmetry realization must be via a PUR
(equivalently by a UR of a suitable central extension) thanks to
Wigner's theorem. {\em If} the centrally extended group happens to be a
`Heisenberg group' i.e. has a unique (up to unitary equivalence) unitary,
irreducible representation, then different quantizations can at most
differ in the multiplicities (reflected in the degeneracies of the
energy spectrum).

It will turn out that the expected symmetry group of the configuration
space (the cylinder), namely the cylinder group has no continuous PURs
and thus there {\em cannot} be {\em any} quantization which will
realize the symmetry. However, we will find that its subgroup, $SO(2)
\times \mathbb{Z}$, can be admitted as the symmetry group and exhibit a
quantum state space which does so together with the dynamics.

The important physical implication is that in this system the electron
will have only the reduced symmetry of discrete axial translations.

The paper is organized is follows.

In section II, we present the analysis based on the PUR's of symmetry
groups. This is divided in five subsections. After describing the
symmetry based approach we recall some basic facts from projective
unitary representations of Abelian groups and their relation to the unitary
representations of their central extensions by $U(1)$ in subsection A.
The approach is illustrated for the familiar planar Landau electron
problem, in subsection B. The non-existence of PUR's of the cylinder
group is discussed next  in subsection C. In subsection D, the PUR's of
the subgroup of the cylinder group, appropriate for discrete axial
translations,  are obtained. In subsection
E, the wave function representation together with the projective action
are detailed. The form of the invariant Hamiltonian is also derived
here. 

In section III, we carry out the analysis beginning with the Hamiltonian
of the system. This is divided in four subsections. In subsection A, we
note the key points which lead to the result and specify the class of 
wavefunctions and the (gauge dependent) Hamiltonian. In subsection B, we
specify the gauge transformation, parameterize the gauge potentials
giving the same magnetic field and classify them into gauge equivalence 
classes using the notion of holonomy of gauge potentials. This is followed 
by the definition of symmetry transformations in this context in subsection C.
Here, the classification of gauge potentials is used to deduce that only 
transformations corresponding to the discrete subgroup of the cylinder group 
qualify to be termed symmetry transformations. The apparent reduction of 
symmetry is manifested at this stage itself. In the last subsection D, the 
spectrum of the Hamiltonian and unitary operators implementing the symmetries 
are presented and shown to reproduce the results from section II.

Section IV contains concluding remarks. 

\section{The Symmetry Approach}
The approach taken in this section can be described as constructing a 
quantum system as realizing certain natural symmetry groups. The approach 
begins by specifying a configuration space for a given system and choosing 
a group, $G$, of transformations of the configuration space as the group of
symmetries of the system. Quite generally, symmetry groups are realized 
in a quantum framework as projective, unitary representations (PUR),
unitary representations being a special (trivial) case of these. This follows 
from the superposition principle via the Wigner theorem \cite{wigner} on
symmetries. Mathematically, such projective representations can be obtained 
as ordinary unitary representations of suitable central extensions $\tilde{G}$
of $G$. The statement that $G$ is a symmetry group of the system entails the
existence of a family of Hamiltonians, one for each equivalence class of
central extensions, and invariant under the action of that particular
extension \cite{divakaran}. 

For example, let the physical system be a charged particle moving in a
plane in the presence of a uniform magnetic field of strength $B$, perpendicular
to the plane. The obvious symmetries of the system form the
Euclidean group, $E_2$, the semi-direct product of the rotation group
$SO(2)$ and the plane translation group, $T$.  Its {\em unitary}
representation in the Hilbert space $L^2(\mathbb{R}^2)$ of square integrable
complex functions on the plane accounts for the free particle case ($B =
0$) while its non-trivial PUR's account for the non-zero magnetic field
case. Since we will use this approach to consider electron on a
cylinder, we will first discuss briefly the usual Landau problem from the
symmetry point of view. It turns out that for the classification of the
PURs of $E_2$, it is sufficient to concentrate on the subgroup of 
translations, $T$.

\subsection{Preliminaries}
\label{Prelim}

Let us briefly recall some basic definitions and facts about PURs and
central extensions of a general Abelian group, $G$. In the following, 
the group multiplication in $G$ is denoted additively.

A PUR of $G$ is a UR up to a phase i.e. if $U$ is a PUR of $G$, 
then $U(g)U(h) = \gamma(g,h) \ U(g + h) \ \forall g, h \in G$, where 
$\gamma(g,h)$ is a 2-cocycle, namely, a complex valued function
with absolute value 1, satisfying certain conditions. If $\gamma$ is of
the form $\gamma(g,h) = \delta(g) \delta(h) \delta(g + h)^{-1}$
with $\delta$ mapping the group elements to complex numbers of modulus
1, then it is a coboundary and the corresponding PUR is a UR of $G$.

PURs of $G$ are more conveniently described in terms of the so-called 
{\em group commutator} of $U(g)$ and $U(h)$,
\begin{equation} \label{Commutator}
%[U(g), U(h)] ~ := ~ 
c(g,h) ~:=~ U(g) U(h) U(g)^{-1} U(h)^{-1} 
~ = ~ \gamma(g, h) \gamma(h, g)^{-1}~~,~ \forall g, h \in G.
\end{equation}

We shall call $c$ the commutator function on $G \times G$ corresponding
to the PUR $U$. The definition of a PUR implies that $c$ satisfies:
\begin{equation} \label{CondComm}
c(g + h, k) ~ = ~ c(g, k) c(h, k) ~,~ 
c(g, h + k) ~ = ~ c(g, h) c(g, k) ~,~ 
c(g, h) c(h, g) ~ = ~ 1 .
\end{equation}

The functions $c$ on $G \times G$ satisfying (\ref{CondComm})
characterize fully equivalence classes of PURs of $G$: the identity function 
$c = 1$ corresponds to the class of trivial PURs, namely the URs of $G$
and the commutator function is unchanged when a 2-cocycle is modified by a 
coboundary. When $G$ is a Lie group and the PUR is a continuous one (the only
case of relevance here) one can write
\begin{equation}\label{Calpha}
c(g, h) = e^{i \alpha(g, h)}
\end{equation}
where $\alpha$ is a real valued, continuous, bilinear, antisymmetric function 
of its arguments. To find all PURs of $G$ one needs to obtain all such 
$\alpha$'s and find continuous maps $U$ from $G$ into unitary operators 
on some Hilbert space $\cal{H}$ satisfying (\ref{Commutator}) for each 
$\alpha$. If there are {\it no} non-zero (mod $2\pi$) $\alpha$'s, then $G$ 
has no non-trivial PURs.

As a general example and one which we will use here, take $G$ to be
the direct product of an Abelian group $A$ and its group of irreducible
characters (its Pontryagin dual), $A^*$ i.e. $G = A \times A^*$. Given
any two characters $\chi, \phi$ of $A$, define
\begin{equation}
c(\ (g, \chi)\ , \ (h, \phi)\ ) := \chi^{-1}(h) \phi(g) ~~~~ \forall ~~ (g,\chi),
(h,\phi) \in A \times A^* .
\end{equation}

It is easy to see that this function defines a commutator function for
$G$. Thus every pair of characters of $A$ defines a class of PURs of 
$A \times A^*$.

Fix a $c = e^{i \alpha}$ corresponding to a PUR of a general Abelian group, 
$G$ and consider the set of ordered pairs $(g, s)$ of elements of $G$ and 
$U(1)$. Define the composition law,
\begin{equation} \label{Composition}
(g,s) \cdot (h,t) := (g + h, e^{i \beta(g, h)} s t)~,~ \mbox{such that} ~
\alpha(g,h) = \beta(g, h) - \beta(h, g) .
\end{equation}

From Eq.(\ref{Commutator}), we can write a cocycle corresponding to $c$
as $\gamma(g,h) = e^{i\beta(g,h)}$. But $\beta$ is not uniquely defined
by (\ref{Composition}). In many (but not all \cite{generaltheory})
cases, we can choose $\beta$ itself to be antisymmetric: $\beta =
\frac{1}{2} \alpha$, which is a canonical choice. In the physical
context, this will be seen to correspond to working in the symmetric gauge. 
We will see in section \ref{Repn} that this canonical choice is impossible 
for the periodic cylinder forcing us to work in a Landau-like gauge.

It follows that this composition makes the set of ordered pairs into a
(non-Abelian) group which is denoted as $G \times_{c} U(1)$ or
$\tilde{G}_c$. This group has $U(1)$ as a central subgroup such that
$\tilde{G}_c / U(1) \sim G$: it is a central extension of $G$ by $U(1)$.
If and only if, $c = 1$ identically, the
corresponding central extension is trivial, i.e. $\sim G \times U(1)$ and 
$G$ is a subgroup of $\tilde{G}_c$. It is easy to verify that the group 
commutator in $\tilde{G}_c$ is given by,
\begin{equation}
(g,s)\cdot(h,t)\cdot(g,s)^{-1}\cdot(h,t)^{-1} ~ = ~ (0, c(g,h)) 
\end{equation}
independent of $s,t$, and so determines a commutator function on $G$.

Now consider a unitary representation $\tilde{U}$ of $\tilde{G}_c$ such
that $\tilde{U}(0, s) = s$. Then it follows that
\begin{equation}
%[\tilde{U}(g,s),\tilde{U}(h,t)] 
\tilde{U}(g,s) \tilde{U}(h,t) \tilde{U}(g,s)^{-1} \tilde{U}(h,t)^{-1}
= c(g,h) .
\end{equation}
In particular, $U(g) := \tilde{U}(g,1)$ is a PUR of $G$ corresponding to the
commutator function $c$. The important result is that {\it every} PUR of $G$ 
arises this way from a UR of one of its central extensions. For further details
we refer the reader to \cite{generaltheory}. One physically significant
consequence of these facts is that the group that acts {\it linearly} 
(unitarily) on the Hilbert space of a system having symmetry group $G$ 
is, in general, {\it not} $G$ itself but some central extension $\tilde{G}$ of
$G$ by $U(1)$. 

\subsection{The Landau electron}
\label{Landau}

We expect the symmetry group of a charged particle moving on the plane with 
a uniform magnetic field $B$, perpendicular to it to be the Euclidean
group, $E_2$. To apply the general procedure outlined above to this
case, we note that central extensions of $E_2$ are completely determined by 
those of its translation subgroup $T$\cite{div-raj}. $T$ is the group of 
two dimensional vectors, $\vec{x}, \vec{y}, \cdots$, under addition. It
is also the configuration space of the particle. A general real, bilinear, 
antisymmetric function of $\vec{x}, \vec{y}$ is clearly of the form:
\begin{equation} \label{LandauCom}
\alpha_{\lambda}(\vec{x}, \vec{y}) = \lambda \vec{x} \wedge \vec{y} 
~ := ~ \lambda (x_1 y_2 - x_2 y_1)
\end{equation}
where $\lambda$ is any real number. It is easy to see that when the group 
commutator defined by the extension $\tilde{T}_{\lambda}$ is specialized to 
group elements close to the identity, one gets the Lie algebra bracket among 
the generators ($P_1, P_2$ forming the basis of the Lie algebra) of 
$\tilde{T}_{\lambda}$ as:
\begin{equation}
[P_1, P_2] = i \lambda \ .
\end{equation}

Note that $\tilde{T}_{\lambda}$ has one extra generator (corresponding to 
the $U(1)$) which commutes with the remaining generators and can be taken 
to be the identity operator in any irreducible UR,
$\tilde{U}$ of interest, since $\tilde{U}(0,s) = s$. This is suppressed in 
the equations. A non-zero $\lambda$ labels equivalence classes of
non-trivial PURs of the translation subgroup while $\lambda = 0$ corresponds 
to the URs. As stated earlier, every central extension $\tilde{T}_{\lambda}$ 
extends to a central extension of the Euclidean group and all central 
extensions of $E_2$ arise this way.
$\tilde{T}_{\lambda \ne 0}$ is the familiar Heisenberg group as is
apparent from the commutator between $P_i$. As is well known, this
has a unique (up to unitary equivalence), continuous, irreducible, infinite 
dimensional, unitary representation. The most familiar concrete form of 
this UR, from the quantum mechanics of a particle in one dimension, is on
$L^2(\mathbb{R})$.

Since our configuration space is $\mathbb{R}^2$, we are interested in the UR of
$\tilde{T}_{\lambda}$ on $L^2(\mathbb{R}^2)$ and this UR, called the wave
function UR,  is not irreducible. Again the general theory 
\cite{heisenberg} implies that this particular representation is 
precisely characterized as the unique, irreducible UR of $\tilde{T}_{\lambda} 
\times \tilde{T}_{-\lambda}$. Equivalently, $L^2(\mathbb{R}^2) ~ \sim ~ 
{\cal{V}}_{\lambda} \otimes {\cal{V}}_{-\lambda}$ where 
${\cal{V}}_{\pm \lambda}$ are the representation spaces of the irreducible 
URs of $\tilde{T}_{\pm \lambda}$. Explicitly, the wave function
representation is obtained as,
\begin{equation} \label{Repren}
\left( W_{\lambda}(\vec{x}, \vec{y}) \psi \right)(\vec{w}) := 
e^{i\frac{\lambda}{2} ( \vec{x} \wedge \vec{w} - \vec{y} \wedge
\vec{w})} \psi(\vec{w} + \vec{x} + \vec{y}) .
\end{equation}
Here $(\vec{x}, 1) \in \mathbb{R}^2 \times_{\lambda} U(1), (\vec{y}, 1) \in
\mathbb{R}^2 \times_{-\lambda} U(1)$ and $\vec{w} \in \mathbb{R}^2$ and 
verification is a matter of simple algebra.

In the present subsection, we have adopted the canonical choice of the
cocycle $\beta = \frac{1}{2}\alpha$. For the real Heisenberg group, this
is always possible. When the identification with the plane Landau
problem is made, a choice of the cocycle is the same as fixing the gauge.
The above canonical choice corresponds to the symmetric gauge. Thus
$\psi$ in (\ref{Repren}) is a symmetric gauge wavefunction.

This completes the description of the kinematics of the projectively
realized translation symmetry of a particle in $\mathbb{R}^2$, namely, the
specification of the state space ${\cal{H}}_{\lambda} = L^2(\mathbb{R}^2)$
and the action of the symmetry group on it via an irreducible UR of
$\tilde{T}_{\lambda} \times \tilde{T}_{-\lambda}$. We emphasize that the
symmetry group gives rise to a central extension $\tilde{T}_{\lambda}$
acting unitarily on ${\cal{V}}_{\lambda}$ while the second factor
$\tilde{T}_{-\lambda}$ keeps track of the multiplicity of the
irreducible UR of $\tilde{T}_{\lambda}$ in the state space ${\cal
H}_{\lambda}$.

To specify dynamics, we need to pick an operator $H_{\lambda}$ on ${\cal
H}_{\lambda}$, the Hamiltonian, which is invariant under the action of
$\tilde{T}_{\lambda}$ (but not under the action of
$\tilde{T}_{-\lambda}$). From the commutation relations between the
momenta $P_i$, it is clear that no polynomial in these can be so invariant
and hence the Hamiltonian {\it cannot} be a polynomial in $P_i$; the
Hamiltonian must be an operator of the form $1 \otimes {\cal{O}}$. We have
however the position operators $X_1, X_2$ on the Hilbert space. Consider
the following combinations (we have set $\hbar = 1$):
\begin{equation}
K_1 := P_1 - \lambda X_2, ~~~~~ K_2 := P_2 + \lambda X_1 .
\end{equation}
These satisfy,
\begin{equation}
[K_1, K_2] = - i \lambda ~,~~~[K_i, X_j] = - i \delta_{ij} ~,~~~
[K_i, P_j] = 0 .
\end{equation}
So $K_i$ form a basis for the Lie algebra of $\tilde{T}_{-\lambda}$ and hence 
map ${\cal{V}}_{-\lambda}$ to itself. Since the actions of $P_i$ and $K_i$ 
commute in ${\cal{H}}_{\lambda}$, a polynomial in $K_i$ is a candidate for a 
Hamiltonian. A simple choice, consistent with the rotational symmetry is:
\begin{equation}
H_{\lambda} := \frac{1}{2 m} \left( K_1^2 + K_2^2 \right).
\end{equation}
This immediately leads to,
\begin{eqnarray}
V_i := \frac{d X_i}{d t} & = & i [ H_{\lambda}, X_i ] ~ = ~ K_i/m  \ , \\
m \frac{d^2 X_i}{d t^2} & = & i [ H_{\lambda}, K_i ] ~ = - \lambda
\epsilon_{ij}K_j \ .
\end{eqnarray}

Thus $K_i$, which is {\em not} the momentum, is actually proportional to
the velocity. Identifying $\lambda = e B$, we see that the Hamiltonian is 
exactly that for a charged particle in a plane with a uniform magnetic field 
perpendicular to the plane. The commutation relations between the
velocity operators, $K_i$, lead to the usual spectrum of the Landau
problem with $\text{dim}({\cal{V}}_{\lambda}) = \infty$ as the degeneracy. 
The usual degeneracy is thus manifestly a consequence of the translation 
symmetry which is however realized {\it projectively}.

We also note here the physical meaning of the commutator function:
\begin{equation}\label{PlaneComm}
\alpha_{\lambda}(\vec{x}, \vec{y}) = e B \vec{x} \wedge \vec{y} =: e
\Phi(\vec{x}, \vec{y}) = 2 \pi \Phi(\vec{x}, \vec{y})/ \Phi_0 \ ,
\end{equation}
where $\Phi(\vec{x}, \vec{y})$ is the flux of $B$ through the
parallelogram of sides $\vec{x}, \vec{y}$ and $\Phi_0 = \frac{2 \pi}{e}$
is the value of the flux quantum.

This demonstrates that the translation symmetric quantum mechanics of a
particle in the plane covers not only free motion ($\lambda = 0$)
but also the Landau electron ($\lambda = e B$). The belief that the
fundamental physical quantity is the field itself - which is
translationally invariant - rather than the vector potential - which is
{\it not} invariant - is thereby vindicated. Gauges and gauge
transformations turn out to be artifacts of the many equivalent ways of
choosing the 2-cocycle and of constructing the unique irreducible URs of 
$\tilde{T}_{\lambda}$ \cite{div-raj} and can be very naturally accommodated. 

Our treatment of the particle on a cylinder from both the representation
theoretic and the `geometric' viewpoint will highlight the care needed
in dealing with a topologically nontrivial configuration space. Once
that is done, both approaches will lead to exactly the same conclusions.

\subsection{Magnetic field on a cylinder}

We now attempt to carry over the representation-theoretic method to the
motion of a charged particle on the surface of an infinite cylinder $S^1
\times \mathbb{R}$ in the presence of a uniform, radial magnetic field, $B$. The
physical realization of such a magnetic field is of no concern to us
here. The group of symmetry transformations, called the cylinder group, is 
$C := SO(2) \times \mathbb{R}$. Topologically, $SO(2)$ is of
course $S^1$. For clarity we denote the circle as a manifold by $S^1$,
as a transformation group by $SO(2)$ and as the multiplicative group of phases
by $U(1)$, as we have already done. The magnetic field of course does not 
break the symmetry under $C$. 

If the corresponding quantum theory can be fully formulated in terms of $B$, 
without invoking a vector potential, then we expect it
to be describable, exactly as before, by continuous PURs of the cylinder
group. If not, then the symmetry under $C$ is necessarily broken.

The striking fact is that the cylinder group has {\it no} continuous,
nontrivial, PURs at all! This follows directly from the general form of
the commutator function, $c(g,h)$, given in (\ref{Calpha}). Denoting a group
element
in $C$ as a pair $(\theta, \eta)$ with $\theta \in SO(2)$ and $\eta \in 
\mathbb{R}$, the commutator functions of $C$ must be of the form $e^{ i 
\lambda (\theta \eta^{\prime} - \theta^{\prime} \eta)}$. The only way this 
can be invariant under the $2 \pi$ periodicity of the angles $\theta, 
\theta^{\prime}$ for all $\eta, \eta^{\prime}$ is if $\lambda$ is zero. 

It is to be stressed that if we were to replace the cylinder group $C$
by its universal covering group $\bar{C} \sim \mathbb{R}^2$, or look only
at its Lie algebra (also $\mathbb{R}^2$), this negative result no longer
holds. Effectively, such a replacement amounts to replacing the
configuration space $S^1 \times \mathbb{R}$ by $\mathbb{R}^2$ and this
is unjustified. Indeed, it is the nontrivial topology of the cylinder group 
which in turn is dictated by the topology of the configuration space that 
is responsible for the non-existence of nontrivial PURs of $C$.
The true symmetry group of the cylinder is $C$ and {\it not} $\bar{C}$. The 
question of when a symmetry group can be replaced by its universal cover for 
quantum mechanical purposes is discussed at length in \cite{divakaran}. 

Physically this means that there is no way of constructing quantum
mechanics of a charged particle on a cylinder in the presence of a
uniform radial magnetic field if the state space is to carry a PUR of
the translation group $C$ of the cylinder. This is a rather strong
conclusion as it does {\it not} depend on any {\it a priori} assumptions
about the details of the function space (with suitable boundary
conditions) to serve as the state space, leave alone the specific choice
of the Hamiltonian as an operator on that space. It does not negate the
value of a formulation of the theory depending essentially on the vector 
potential, but we should not then expect to fully implement the translation 
invariance. In the next section a treatment based on vector potential will 
be presented with essentially the same conclusion, that the full translation 
symmetry is not possible.

We have already noted that infinitesimal translations, encoded in the Lie
algebra of $C$, and hence {\it not} distinguishing between the cylinder
and the plane, do have non-trivial PURs. This has a useful consequence. We can
``integrate" a PUR of $C$ along  any curve which does not fully wind around
the cylinder. More precisely, consider a contractible region $D$ of $S^1
\times \mathbb{R}$. Suppose $D$ contains the `rectangle' with vertices
at $(0, 0), (\theta, 0), (\theta, \eta)$ and $(0, \eta)$. Then from the
Lie algebra central extension
\begin{equation}
[P_{\theta}, P_{\eta}] = i \lambda ~,
\end{equation}
we have the group commutator,
\begin{equation}
e^{i \theta P_{\theta}} e^{i \eta P_{\eta}}
e^{-i \theta P_{\theta}} e^{-i \eta P_{\eta}}
= e^{i \lambda \theta \eta} ~.
\end{equation}
If we apply a uniform radial magnetic field to a particle moving in $D$,
then its momentum will satisfy these equations with the identification
$\lambda = e B$, exactly as in the case of the plane
eq.(\ref{PlaneComm}). Hence, locally within $D$, we have the old relationship
\begin{equation} \label{flux}
e^{i \theta P_{\theta}} e^{i \eta P_{\eta}}
e^{-i \theta P_{\theta}} e^{-i \eta P_{\eta}}
= e^{i e B \theta \eta}
~~ =: ~~ e^{2 \pi i \frac{\Phi(\theta, \eta)}{\Phi_0}}
\end{equation}
where $\Phi(\theta, \eta)$ is the flux through the rectangle and
$\Phi_0$ is the flux quantum. This will break down as soon as $D$ (and
the rectangle in it) becomes non-contractible.

\subsection{The Periodic cylinder}

While the cylinder group has no nontrivial PURs, it has a subgroup which
does. One can apply our methods to specify the state space and the
dynamics of the Landau electron having the reduced symmetry of the
subgroup. 

The key fact is that the subgroup $C_{\mathbb Z} := SO(2) \times \mathbb{Z}$ 
of $C$, has the structure $A \times A^*$. This is because the
group of characters of $SO(2)$ is ${\mathbb Z}$ and vice versa, i.e.
$SO(2)^* = {\mathbb Z}$
and ${\mathbb Z}^* = SO(2)$. From the general theory summarized in
\ref{Prelim}, $C_{\mathbb Z}$ has a Heisenberg central extension given by 
the commutator function,
\begin{equation} \label{ComFun}
c_{\nu}( ~ (\phi, m), (\phi^{\prime}, m^{\prime}) ~ ) := e^{i \nu
(m^{\prime} \phi - m \phi^{\prime}) }, 
\end{equation}
where $\nu$ is an integer since $\phi$ is an angle.

Physically one may imagine reducing $C$ to $C_{\mathbb Z}$ by applying a
periodic potential $V$, along the axial direction: $V$ is independent of
$\phi$ and $V(\eta + \ell) = V(\eta)$ for a fixed $\ell$ which we take
to be 1. Such a potential will have to be added to the Hamiltonian, but it 
can be taken to be arbitrarily small. Its only purpose here is to reduce the 
symmetry group. 

In looking for the quantum mechanical realization of the PURs of
$C_{\mathbb{Z}}$ through a magnetic field $B$, it is to be expected that
there may be restrictions on the possible values of $B$, arising from
the quantization of the central charge $\nu$. This in fact is the case.
From eq.(\ref{ComFun}), we have the condition
\begin{equation}
c_{\nu}\left( (2 \pi, 0) , (0, 1) \right) = 1
\end{equation}
for the commutator of a translation through a period along the axis and
by $2 \pi$ around the axis implying that $\nu$ is an integer. But from the 
eq.(\ref{flux}), in the limit $\phi \to 2\pi$ and $\eta = 1$, we also have 
\begin{equation}
c_{\nu}\left( (2 \pi, 0) , (0, 1) \right) = e^{i 2\pi
\frac{\Phi}{\Phi_0}}
\end{equation}
where $\Phi$ is the flux through a unit slice of the cylinder. We
conclude:

{\em The periodic cylinder group $C_{\mathbb{Z}} = SO(2) \times
\mathbb{Z}$ has a family of PURs labelled by integers, describing at least at 
the kinematical level, the quantum mechanics of a charged particle in a 
uniform radial magnetic field $B$ if and only if $B$ is such that the flux 
per slice is an integral multiple of the flux quantum.}

In this form the conclusion is dimension-free: it is independent of the
radius of the cylinder and the period of the periodic potential. The
corresponding central extensions $C_{\mathbb{Z}} \times_{\nu} U(1)$ are 
classified by the quantized values of the flux/slice. 
%The group composition law is:
%%
%\begin{equation}
%\left( (\theta, m, 1)~,~(\theta^{\prime}, m^{\prime}, 1)\right) = 
%\left( \theta + \theta^{\prime} (mod 2 \pi)~,~ m + m^{\prime}, e^{i
%\frac{m^{\prime} \theta - m \theta^{\prime}}{2 \Phi_0}} \right) .
%\end{equation}
%
The only `magnetic quantity' that enters in the description is the
flux/slice and not $B$ itself.

\subsection{Representations, wave functions, dynamics}
\label{Repn}

It is convenient to borrow from the mathematics literature the term
``Heisenberg group" to designate a nontrivial central extension by
$U(1)$ of an Abelian group $G$ having the structure $A \times A^*$
\cite{heisenberg}. The distinguishing feature of a Heisenberg group is
the Stone-von Neumann property: all its continuous irreducible unitary
representations are unitarily equivalent. The essentially unique
irreducible UR can be concretely realized on $L^2$ functions on certain
quotient groups of $G$ (all of which are equivalent), in particular, on
$L^2(A)$ and on $L^2(A^*)$. The ``original" Heisenberg group
$\mathbb{R}^2 \times_{\lambda} U(1), \lambda \in \mathbb{R}$ provides a
familiar illustration of all these properties.

The centrally extended periodic cylinder group
$\tilde{C}_{\mathbb{Z}\nu} := ( SO(2) \times \mathbb{Z} ) \times_{\nu}
U(1)$ is a Heisenberg group and has equivalent irreducible URs on
$L^2(S^1)$ and $L^2(\mathbb{Z})$. 

On $L^2(S^1)$, the action of $\tilde{C}_{\mathbb{Z}\nu}$ is given by
\begin{eqnarray}
\left( U_{S^1} (\theta, 0) f \right)(\phi) & := & f(\theta + \phi)
\nonumber \\
\left( U_{S^1} (0, m) f \right)(\phi) & := & c_{\nu}\left( (\phi,0),
(0,m)\right) f(\phi) ~ = ~ e^{i \nu m \phi} f(\phi).
\end{eqnarray}

Likewise, on $L^2(\mathbb{Z})$, the action of $\tilde{C}_{\mathbb{Z}\nu}$ is 
given by
\begin{eqnarray}
\left( U_{\mathbb{Z}} (\theta, 0) f \right)(n) & := & 
c_{\nu}\left( (0, n), (\theta, 0)\right) f(n) ~ = ~ e^{-i \nu n \theta} \ , \nonumber \\
\left( U_{\mathbb{Z}} (0, m) f \right)(n) & := & f(n + m) \ .
\end{eqnarray}

For the wave function representation, we look for representations of the
symmetry group $\tilde{C}_{\mathbb{Z}\nu}$ on square integrable
functions, or more generally on $L^2$ sections of a line bundle, on
the configuration space $S^1 \times \mathbb{R}$. These are characterized
by a quasi-periodicity parameter $q$ defined as $\psi(\theta + 2 \pi, y) =
e^{2 \pi i q} \psi(\theta, y)$ and the Hilbert space is denoted by
$ {\cal{H}}_{\nu} := L^2_q(S^1 \times \mathbb{R})$. The UR $W_{\nu}$ of $\tilde{C}_{\mathbb{Z}\nu}$
is explicitly defined for a fixed $q$ as:
\begin{equation} \label{UCz}
(\ W_{\nu}(\phi, m) \psi\ )(\theta, y) := e^{i (\nu m \theta - q \phi)}
\psi(\theta + \phi, y + m) \ .
\end{equation}

The phase factor in (\ref{UCz}) corresponds to a choice of cocycle which
is {\em not} the canonical one. The reason is that the square root of
the commutator function is not well defined on $C$. It is easily
checked that this choice is compatible with the commutator function in
(\ref{ComFun}). One can also see that $W_{\nu}(2 \pi, 0) = \text{Id}$ on
the state space ${\cal{H}}_{\nu}$. This representation of 
$\tilde{C}_{\mathbb{Z}\nu}$ cannot be irreducible.

Remarkably, this space ${\cal{H}}_{\nu}$ also carries a UR, $V_{-\nu}$,
of the real Heisenberg group corresponding to central charge $-\nu$.
This is defined explicitly as:
\begin{equation} \label{UHeis}
(\ V_{-\nu}(\xi, \eta) \psi\ )(\theta, y) := e^{i \nu \xi y }
\psi(\theta + \xi, y + \eta) \ .
\end{equation}

Furthermore, the actions of the two groups commute. The phase factor in 
(\ref{UHeis}) corresponds to a choice of cocycle which is also {\em not} 
the canonical one. It is for this choice that the two group actions commute. 
Once again, the commutator function given in (\ref{LandauCom}) is  unchanged.
These facts are easily checked.

It follows from this that state space has the decomposition:
\begin{equation}\label{HDecomp}
{\cal{H}}_{\nu} = {\cal{W}}_{\nu} \otimes {\cal{V}}_{-\nu} ,
\end{equation}
where ${\cal{W}}_{\nu}$ and ${\cal{V}}_{-\nu}$ are representation
spaces for $W_{\nu}$ and $V_{-\nu}$ respectively.

For every fixed $q$, one also has the decomposition, $L^2_q(S^1 \times
\mathbb{R}) = L^2_q(S^1) \otimes L^2(\mathbb{R})$. This fact
together with eq.(\ref{HDecomp}) implies that ${\cal{H}}_{\nu}$ is an
irreducible UR of $\tilde{C}_{\mathbb{Z}\nu} \times
\tilde{\mathbb{R}}^2_{-\nu}$, since each factor in the tensor product is
irreducible under the corresponding subgroup.

We emphasize that the first factor in the tensor product decomposition
of the state space represents the symmetry group while the second factor
represents the multiplicity of the representation. The second
factor is where the dynamics will be defined.

As before, the only operators invariant under the symmetry group 
$\tilde{C}_{\mathbb{Z}\nu}$ are of the form $1 \otimes {\cal{O}}$ and the 
correct invariant Hamiltonian is:
\begin{equation}
H_{\nu} = 1 \otimes \frac{( K_1^2 + K_2^2 )}{2 m}
\end{equation}
with $[K_1, K_2] = - i \nu = -i e B_{\nu}$ and $B_{\nu}$ such that the
flux $2 \pi B_{\nu}  = \nu \Phi_0$. Again, as before, one verifies that $K_i$
are indeed the velocity operators. Strictly speaking we also have to add
a periodic potential needed to reduce the symmetry group. The strength
of this potential can be taken to be arbitrarily small so that it defines 
an ``empty lattice" along the axis of the cylinder.

In the empty lattice limit, the spectrum of the Hamiltonian is exactly
as in the plane. This is not surprising since velocities are tangent to
the cylinder and so are insensitive to whether one is working on the
plane or the cylinder. The one difference is that not only is the energy
quantized as a function of the magnetic field but the flux itself can
only take allowed quantized values for which PURs of
$C_{\mathbb{Z}}$ exist. Every energy eigenvalue has multiplicity 1 in 
${\cal{V}}_{-\nu}$ while ${\cal{W}}_{\nu}$ accounts for the infinite
degeneracy of the energy eigenvalues in the full state space, exactly as
in the plane problem.

\section{The Hamiltonian Approach}

In the previous section we saw the construction of the quantum mechanics
(state space and dynamics) of a charged particle in a uniform magnetic field
first on a plane and then on a cylinder. The distinctive feature of that 
approach was that a {\em symmetry group} was chosen {\em a priori}, as a group
of transformations of the configuration space of the system under which the 
specification of the system was invariant, namely the Euclidean group in the 
case of the plane and the cylinder group, $SO(2) \times \mathbb{R}$, in the 
case of the cylinder. Although the system was `obviously' invariant under 
these groups, surprisingly at the quantum level the symmetry was reduced to 
$SO(2) \times \mathbb{Z}$ in the case of the cylinder while in the case of 
the plane, there was no such reduction of the symmetry. This followed
from the non-existence of non-trivial, projective unitary representations of 
the cylinder group. In this analysis, the Hamiltonian played {\em no} explicit
role and was in fact derived by demanding invariance under the symmetry 
action. There was no explicit introduction of a gauge potential, and the 
approach was manifestly gauge invariant. 

In this section we will arrive at the same conclusions using the more familiar
approach in which the Hamiltonian is chosen to begin with and the
corresponding admissible symmetries are deduced as a consequence. This
will also serve to clarify the surprising nature of the result. In brief, it 
will turn out that, in a certain sense, the full cylinder symmetry was never 
there to begin with! We will concentrate only on the cylinder case with 
reference to the plane only for comparison. Throughout this section,
{\em physical units} are used in order to highlight the points at which 
quantum considerations come into play. 

\subsection{Key points and the basic set up}
\label{key}

It is worth keeping in mind the following three points to appreciate the
steps of the arguments leading to the result. These points are
elaborated further subsequently.

(1) Since in the present approach, we begin with an a priori chosen classical
Hamiltonian corresponding to the interaction of a charged particle with
an external magnetic field, we will have to introduce a gauge potential to be 
able to write the Hamiltonian in a local form. The same of course is also true 
in a formulation based on an action principle. One then has the freedom to 
choose any of the gauge potentials which give the same magnetic field. While 
stipulating the class of transformations of the cylinder which qualify to be 
symmetries within the formulation, one has to pay attention to this freedom. 

(2) The topology of the cylinder implies that there are several
gauge potentials, giving the same magnetic field, which however do
{\em not} differ by the gradient of functions on the cylinder. These 
are classified further by the notion of `holonomy equivalence'. It will turn 
out that there are infinitely many holonomy equivalence classes on the
cylinder while there is only one such class on the plane. A complete
gauge invariant specification of the system is made by giving a single
holonomy equivalence class. 

(3) At the quantum mechanical level, in the context of a Schr\"odinger
like representation, the wave functions (sections of line bundles) have to 
be chosen to be quasi-periodic. As usual, functions of the coordinates
will act by multiplication. To preserve the quasi-periodicity of the wave 
functions, such multiplicative operators must be single valued. In particular
the gauge potentials must be single valued. 

The cylinder is taken to be coordinatized in the usual manner by the
angular coordinate $\theta$ and the coordinate $y$ along the axis. The
corresponding conjugate momenta are $p_{\theta}, p_y$ and the radius of 
the cylinder is $R$. The classical Hamiltonian is given by (we have chosen the
units so that the speed of light in vacuum is 1),
\begin{equation}
H( \theta, y, p_{\theta}, p_y; A(\theta, y)) := \frac{(p_y - e
A_y(\theta, y))^2}{2 m} +
\frac{(p_{\theta} - e A_{\theta}(\theta, y))^2}{2 m} R^{-2}
\end{equation}

Quantum mechanically, this is to be represented by a self adjoint operator
on a suitable Hilbert space, with the conjugate momenta acting as derivative 
operators and coordinates acting as multiplication operators. Explicitly, 
$p_y := -i \hbar \partial_{y} \ , p_{\theta} := -i \hbar \partial_{\theta}$. 
To ensure that the momenta, in particular $p_{\theta}$, are self adjoint,
their domain must consist of wave functions $\psi(\theta, y)$ which are
in general quasi-periodic: $\psi(\theta + 2 \pi, y) = e^{2 \pi i q}
\psi(\theta, y)$, where $q$ is a real parameter in the interval $[0,1)$, 
specifying the quasi-periodicity. For the Hamiltonian to be a well defined 
operator on such quasi-periodic functions then, the gauge potential must be 
single valued as already remarked.

\subsection{Classification of gauge potentials}
\label{GaugeClass}

Let us now turn to gauge transformations. Recall that the usual gauge
transformations of the gauge potentials and the {\em phase} transformations 
of the quantum wave function are correlated in the following manner.
Using the notation, $g := e^{i \frac{e}{\hbar}\Lambda}$,
\begin{eqnarray}
\psi^{\prime} := g \psi & ~~ ,
~~ & A^{\prime}_j := g A_j g^{-1} + i \frac{\hbar}{e} g \partial_j g^{-1} ~~~~ 
\Longleftrightarrow ~~ \nonumber \\
\left\{-i \hbar \partial_j  -e A^{\prime}_j \right \}\psi^{\prime} & = &
g \left\{-i \hbar \partial_j  -e A_j \right\}\psi 
\end{eqnarray}

The phase transformation $g$, also has to be single valued to preserve the
quasi-periodicity of the wave functions. This also preserves the
single valued nature of the gauge potential. Notice that if $\Lambda$ is
single valued, so is $g$ but the converse is not true. For example
$g_m(\theta, y) := e^{i m \theta} \ , \ m \in \mathbb{Z}$, is single valued
although $\Lambda \sim \theta$ is not. Incidentally, these are the only
phase transformations which are single valued without the corresponding
exponent $i\frac{e}{\hbar} \Lambda$ being so.

It is convenient to use the differential form notation: 
$A := A_y dy + A_{\theta} d\theta \ , \ d := d y \partial_y + d \theta
\partial_{\theta} \ , \ dA = ( \partial_{\theta} A_y - \partial_y A_{\theta} ) 
d\theta \wedge d y $ etc. In the present context of a uniform radial magnetic 
field of magnitude $B$, we have $d A = B R d \theta \wedge d y$. 

In terms of this notation, two gauge potential 1-forms, $A' , A$, are defined 
to be {\em gauge equivalent} if they are related by a {\em gauge
transformation},
\begin{equation} \label{GaugeTrans}
A^{\prime} ~ = ~ A^g := g A g^{-1} + i \frac{\hbar}{e} g d g^{-1} ~ = ~ A + i
\frac{\hbar}{e} g d g^{-1} 
\end{equation}
for {\em some smooth, single valued function $g$ on the cylinder}. It is
immediate that the corresponding magnetic field 2-forms, $d A', \ d A$ are
equal for all gauge equivalent potentials, expressing the gauge
invariance of the magnetic field. Note that the converse need not be true 
i.e. we may have two gauge potentials giving the same magnetic field which
are nonetheless {\em not} gauge equivalent. 

To see this, observe that the difference of any two gauge potentials giving 
the same magnetic field must be a closed 1-form. On the plane, it being
contractible, every closed 1-form is also exact i.e. it is the exterior
derivative (or gradient) of some function $\Lambda$ on the cylinder. It is 
obvious then that the two potentials are related by a gauge transformation. 
There is thus exactly one gauge equivalence class of the potential and it is 
characterized by the magnetic field. The situation on the cylinder is different.
Every closed 1-form is {\em not} exact. Since the cylinder is simple enough we 
can write down the most general form of a gauge potential:
\begin{equation}
A^{\prime} ~ = ~ A + \zeta d \theta + d \Lambda 
~~~~~~ \Longleftrightarrow ~~~~~~
A^{\prime}_{\theta} ~ = ~ A_{\theta} + \partial_{\theta} \Lambda + \zeta
~~,~~
A^{\prime}_{y} ~ = ~ A_{y} + \partial_{y} \Lambda 
\end{equation}

Here $\zeta$ is a real {\em constant} and $\Lambda$ is of course a function
on the cylinder which by definition is single valued. A convenient 
parameterization of {\em all} the gauge potentials giving the {\em same magnetic
field} is then obtained by choosing 
$A_y = 0 \ , \ A_{\theta} = - B y$ so that,
\begin{equation} \label{GaugeParam}
A^{\zeta} = (\zeta - B R y) d \theta + d \Lambda, ~~~~~ d A^{\zeta} =
B~ (R\, d \theta \wedge d y).
\end{equation}

Note that $A^{\zeta}$ and $A^{\zeta^{\prime} \ne \zeta}$ are {\it not} gauge 
equivalent on the cylinder since $\theta$ is {\it not} a single valued
function (or a $0$-form) on the cylinder. Incidentally single valuedness of
gauge transformation also implies that unlike the planar case, a `symmetric'
gauge for the gauge potential ($ A_y = \frac{B}{2} R \theta, A_{\theta} = 
- \frac{B}{2} R y$) is {\em not} possible. At this stage we have a 1-parameter
family of gauge equivalence classes of the potentials, all giving the same 
magnetic field. Clearly, unlike on the plane, the magnetic field alone does 
{\em not} constitute a {\em complete} characterization of a gauge equivalence 
class.

However, it is well known that there is another way to classify gauge
potentials using their {\em holonomies} (also known as exponentials of 
non-integrable phase factors \cite{yang}) along all closed curves (or
loops for short). Since on a cylinder we do have {\em non-contractible} 
loops, those which wind around the cylinder, we do have a `larger' class of
holonomies compared to the planar case. 

The holonomy of a gauge potential (in the Abelian case) along a 
loop $\tau$ is defined to be:
\begin{equation} \label{HolonomyDef}
h_{\tau}(A) := exp \left( i \frac{e}{\hbar} \oint_{\tau} A \right)
\end{equation}

One can now define a {\em holonomy equivalence} relation on the set of
gauge potentials. Two gauge potentials are said to be {\em holonomically
equivalent} if their holonomies along {\em all} loops are equal.
In the present context we will see explicitly that holonomy equivalence 
implies gauge equivalence and conversely.

Using the parameterization of the gauge potential (\ref{GaugeParam}), it is 
easy to see that 
\begin{eqnarray} \label{HolonomyValue}
h^{\zeta}_{\tau} & = & exp \left\{
i \ \frac{e} {\hbar} \oint \zeta d \theta +
i \frac{e}{\hbar} (- B R) \oint y d \theta  + 
i \frac{e}{\hbar} \oint d \Lambda \right\}
\nonumber \\
& = & exp \left\{
i \ \frac{e} {\hbar} \zeta ~ 2 \pi w_{\tau} \right\} 
~ exp \left\{
- i \ \frac{e} {\hbar} \Phi_{\tau} \right\} \ ,
\end{eqnarray}
where $w_{\tau}$ is the number of times the loop $\tau$ winds around the
cylinder and $w_{\tau} = 0$ corresponds to a contractible loop. For a
contractible loop, the second factor contains the flux of the magnetic
field through the loop. The first factor contains the dependence on the
parameter $\zeta$ while the second one is common to all gauge potentials of
eq.(\ref{GaugeParam}) and is manifestly gauge invariant.

It follows immediately that
\begin{eqnarray} \label{HolonomyCaln}
h^{\zeta^{\prime}}_{\tau} & = & 
h^{\zeta}_{\tau} ~~ exp \left\{
i \ \frac{e} {\hbar} (\zeta^{\prime} - \zeta) ~ 2 \pi w_{\tau} \right\}
\ .
\end{eqnarray}
Clearly, for all contractible loops the holonomies of all $A^{\zeta}$
are equal. For non-contractible loops though the holonomies are equal
if and only if 
\begin{equation} \label{GammaEqui}
\zeta^{\prime} ~ = ~ \zeta + m \left( \frac{\hbar}{e} \right) ~~ , ~~ 
m \in \mathbb{Z} .
\end{equation}

Now, if two gauge potentials are gauge equivalent, their difference is
proportional to $g d g^{-1}$ which is either an exact 1-form or is
proportional to $d \theta$. In either case, their holonomies along all
loops are equal. Thus gauge equivalence implies holonomy
equivalence. To prove the converse, we use the equation
(\ref{GammaEqui}) and the parameterization in (\ref{GaugeParam}). It then
follows that we can first choose a $g_m = e^{i m \theta}$ to make the
two $\zeta$'s equal and then choose a $\Lambda$ to make the two
potentials equal. Thus we see that holonomy equivalence also implies
gauge equivalence. From now on we will refer to them simply as gauge 
equivalence classes and denote them as $[A^{\zeta}]$: 
\begin{equation} \label{EqClassDef}
[A^{\zeta}] := \{ A^{\zeta'} \ / \  A^{\zeta'} = 
(\zeta' -B R y) d \theta + d \Lambda \ , \ \zeta' = \zeta + m
\frac{\hbar}{e}\ , \ m \in \mathbb{Z} \}
\end{equation}

The gauge equivalence classes of gauge potentials are thus labeled by a real
number in the interval $[0, \frac{\hbar}{e})$; shifting $\zeta$ by an integer
multiple of $\frac{\hbar}{e}$ keeps the gauge potential within the same gauge 
equivalence class. The holonomies are functions of the equivalence classes.

\subsection{Definition of symmetry transformations}
\label{SymmDef}

Having specified the quasi-periodicity properties of the wave functions, 
the action of gauge transformations on them and classified the gauge 
potentials, we now turn to the specification of the symmetry transformations.

The classical equations of motion are of course manifestly gauge invariant. 
However, due to the explicit presence of the gauge potential, the Hamiltonian 
is manifestly gauge {\em non-invariant}. In the usual context of ordinary
potentials, symmetry transformations are required to leave the Hamiltonian 
invariant. In the present context, we must allow the Hamiltonian to change 
while formulating the definition of a symmetry transformation. The allowed 
change in the Hamiltonian is not arbitrary but must be such that it can be 
compensated by a gauge transformation of the gauge potential. 

Thus we define {\em symmetry transformations as those transformations of 
the configuration space which induce a gauge transformation on the gauge
potential}. 

This is physically reasonable, since gauge transformations will keep one 
in the same gauge-equivalence class of the potential signifying that the 
definition of symmetry is applied to gauge invariant specification of the 
system. If we insisted on the strict invariance of the Hamiltonian,
then, on the plane also, there is {\em no} full translation symmetry when 
the potential is chosen in any fixed gauge: the identification of a symmetry
will be gauge dependent. 

Let us now focus in particular on the translation symmetry along the $y$ 
direction. Under $\theta^{\prime} = \theta ~,~ y^{\prime} = y + \ell$, the 
gauge potential, being a 1-form, transforms as,
\begin{equation}
A_i^{\prime}( \theta^{\prime}, y^{\prime} ) ~ = ~ \frac{\partial x^j}{\partial
x^{\prime \, i}} ~ A_j( \theta, y)  ~~~ \Longleftrightarrow ~~~
A^{\prime}_i (\theta, y) ~ = ~ A_i ( \theta, y - \ell) \ .
\end{equation}
Applying this to $A = A^{\zeta}$, one sees that 
\begin{eqnarray} \label{ATrans}
(A^{\zeta})^{\prime}(\theta, y) & = & A^{\zeta}(\theta, y - \ell) ~
\Longrightarrow \nonumber \\
\left[(A^{\zeta})'\right] & = & \left[A^{\zeta + B R \ell}\right]
\end{eqnarray}

Thus a translation along the $y$ direction, changes the gauge equivalence 
class label from $\zeta$ to $\zeta + B R \ell$.  It is obvious now that
{\em unless $\ell$ is an integer multiple of $\frac{\hbar}{e B R}$, 
translations along the y-direction are not a symmetry of the system}. 

To summarize: the existence of gauge inequivalent potentials obstructs the
continuous y-translation symmetry but within each gauge equivalence class
there is a reduced symmetry of discrete y-translations with the step
size given by $\frac{\hbar}{e B R}$. The loss of continuous 
symmetry is attributable to the non-availability of sufficiently many single
valued gauge transformation ($\Lambda(\theta, y) \sim \theta$ not allowed) 
while the survival of discrete y-translations is because of the availability
of gauge transformations $g(\theta, y) \sim e^{i m \theta}$.

Several remarks are in order.

(1) Notice that arbitrary y-translations {\em change} the gauge equivalence
classes {\em only} for non-zero magnetic field. For the zero magnetic
field therefore continuous y-translations {\em are} symmetries. In view
of the fact that on non-contractible spaces the magnetic field alone
does not constitute a complete gauge invariant specification but the 
holonomies do, it is important to note that the `free particle' on a
cylinder is to be defined by requiring all holonomies to be equal to 1.
This in particular implies that the magnetic field must be zero {\em
and} the parameter $\zeta$ must also be zero so that the gauge potential
is in the equivalence class of the `zero' gauge potential. In this case
of course the full symmetry group of the cylinder will be realized. We
will deal only with non-zero $B$.

(2) All of the discussion of the classification of (single valued) gauge 
potentials and symmetries can be carried out at the purely classical level. 
There is a point about dimensions to be noted though. Mathematically, a gauge
potential comes from a connection on a principal bundle and its holonomies lead 
to the holonomy equivalence classes. Its dimensions can be chosen to be 
$[A^{\text{mathematical}}] \sim L^{-1}$ so that the exponent in the definition 
of the holonomy is dimensionless. In the mathematical definition of holonomy 
there is of course no $\frac{e}{c \hbar}$. One can see easily that one still 
gets an equation similar to (\ref{GammaEqui}). The subsequent analysis of 
y-translations leads to the same conclusion about the loss of continuous 
y-translation symmetry. In this way one can argue that even at the 
{\em classical} level, there is a loss of continuous y-translation symmetry. 
One will still have the discrete translations as a symmetry, but the step size
will not involve any $\hbar$. In conventional (Gaussian for electromagnetism)
units, $\frac{e}{c}A^{\text{physical}}$ has dimensions of $M^1 L^1 T^{-1}$. To 
define a holonomy of $A^{\text{physical}}$, definable for arbitrary 
configurations of sources, one must use a universal constant with dimensions 
$M^1 L^2 T^{-1}$. There is only one such fundamental constant which is 
$\hbar$. Thus one identifies: $A^{\text{mathematical}} := \frac{e}{c \hbar} 
A^{\text{physical}}$. One sees that in order that a physical vector potential 
exhibits the geometrical properties of a connection on a principal bundle, a 
quantum framework is essential. The step size thus has a quantum origin. 

(3) Note that at the classical level, one could conceivably allow non-single
valued gauge potentials (since the potential plays only a local role).
This will then also enlarge the class of gauge transformations. With
this extended notion of gauge equivalence, one would obtain a single
equivalence class of gauge potentials and consequently continuous
y-translations would be admissible symmetry transformations. Quantum
mechanically, we do {\em not} have this option.

From now on we introduce $\mu := \frac{e B R}{\hbar} =
\frac{\tilde{\Phi}}{\Phi_0}$ which is the magnetic flux {\em per unit length}
of the cylinder in units of the flux quantum. The step size for discrete
y-translations is then denoted by $\mu^{-1}$. In the light of the remark
(1) above, we will also put $\zeta = B R \rho$ for notational convenience 
and use $\rho$ to label the gauge equivalence classes. 

\subsection{Spectrum, wavefunctions and symmetry realization}
\label{SpectSym}

We now proceed to obtain, as in the case of the plane, the spectrum of states, 
their degeneracies and organization in terms of the symmetry group, $SO(2) 
\times \mathbb{Z}$. We will also see that the the symmetry group is
represented projectively.

Let us choose any one of the gauge equivalence class, labeled by 
$\rho$, and write the corresponding Hamiltonian in the usual manner.
Introducing the notation:
\begin{eqnarray}
\pi^{\rho}_y  :=  p_y - e A^{\rho}_y ~~~~~ p_y := -i \hbar
\partial_y &~~~~~,~~~~~&
\pi^{\rho}_{\theta}  :=  p_{\theta} - e A^{\rho}_{\theta} ~~~~~
p_{\theta} := -i \hbar \partial_{\theta} \ ,
\end{eqnarray}
it follows that
\begin{equation}
\left[ \pi^{\rho}_{\theta} , \pi^{\rho}_y \right]  =  i \hbar e
(\partial_{\theta} A^{\rho}_y - \partial_y A^{\rho}_{\theta} ) ~~~~~ 
= ~~ i \hbar e B R
\end{equation}
The Hamiltonian is given by,
\begin{equation}
H^{\rho}  =  \frac{(\pi^{\rho}_y)^2 + R^{-2}(\pi^{\rho}_{\theta})^2}
{2 m} ~~~=~~~ \frac{eB}{2m}( P_{\rho}^2 + Q_{\rho}^2 ), ~~~~~~ 
Q_{\rho} := \frac{\pi^{\rho}_{\theta}}{R \, \sqrt{ e B}} ~,~ P_{\rho}
:= \frac{\pi^{\rho}_y}{\sqrt{ e B }}
\end{equation}

From these, it follows that the energy eigenvalues are $\hbar
\omega ( N + \frac{1}{2} )$ with $\omega := \frac{e B}{\hbar m}$ and
$N$, a non-negative integer.  The eigenvalues are manifestly in terms of the 
magnetic field and are also independent of $\rho$ labeling the gauge 
equivalence class. The degeneracy of all the eigenvalues is the same and
is determined from the degeneracy of the ground state. 

The subspace of ground states is obtained by solving $(Q_{\rho} + i P_{\rho})
\Omega_ {\rho}(\theta, y) = 0$. The explicit, normalized solutions for the 
ground states are given by,
\begin{eqnarray}
\Omega_{\rho, n} (\theta, y) & = &
%\frac{1}{\sqrt{2 \pi}} \left(\frac{\sigma}{\pi}\right)^{\frac{1}{4}} e^{i (n + q) \theta}
%e^{ -\frac{\sigma}{2} (y - \frac{\rho_n}{\sigma})^2 }\nonumber \\
%& = & 
\frac{1}{\sqrt{2 \pi}} \left(\frac{\mu}{\pi R}\right)^{\frac{1}{4}} e^{i (n + q) \theta}
e^{ -\frac{\mu}{2 R} (y - \rho + \frac{n + q}{\mu})^2 }, ~~~~~~
\mbox{for} ~~ n \in \mathbb{Z}. \label{GroundState}
\end{eqnarray}

For a fixed $q$ and $\rho$ one does have infinite degeneracy labeled by
$n$, just as in the case of the plane. Can we organize this degeneracy as 
a representation of the symmetry group $SO(2) \times \mathbb{Z}$? 

To check this we proceed as in the case of the plane. We look for
operators of the form
\begin{eqnarray}
\Sigma^{\rho}_{\theta} := \pi^{\rho}_{\theta} + \sigma_{\theta}(\theta,
y) \ , \nonumber \\
\Sigma^{\rho}_{y} := \pi^{\rho}_{y} + \sigma_y(\theta, y) \ ,
\end{eqnarray}
and require that these commute with the $\pi^{\rho}_i$'s. A quick
calculation shows that $\sigma_{\theta}(\theta, y) = -e B R y +
C_{\theta} $ and $ \sigma_y(\theta, y) = e B R \theta +  C_y $, where
$C_{\theta}, C_y$ are constants. The functions $\sigma_i$ must be
single valued to preserve the quasi-periodicity of the wave functions.
However $\sigma_y$ is not a single valued function on the cylinder and 
consequently $\Sigma^{\rho}_{y}$ is not well defined as an operator. 
One thus has only {\it one} self adjoint generator
$\Sigma^{\rho}_{\theta} = - i \hbar \partial_{\theta} - e B R \rho +
C_{\theta}$ commuting with the Hamiltonian and consequently, only a one 
(continuous) parameter unitary group of symmetries is possible. It generates
translations along the $\theta$-direction. Infinitesimal translations along 
the $y$-direction are {\it not} symmetries of the Hamiltonian for any choice
of the gauge equivalence class. This manifests the loss of continuous 
y-translation symmetry.

But although the function $\sigma_y(\theta, y) = eBR\theta + C_y$ is not well
defined on the cylinder, its exponential, $e^{i \frac{a}{\hbar} \sigma_y}$ 
{\it is}, {\it provided} $a \mu \in \mathbb{Z}$. Therefore, {\it finite} 
translations are admissible and the symmetry group is $SO(2) \times \mathbb{Z}$,
possibly projectively realized. 

Define finite translations operators for $\phi \in [0, 2 \pi)$ and $a
\mu \in \mathbb{Z}$: 
\begin{eqnarray}
U^{\rho}(\phi) ~ := ~ e^{i \frac{\phi}{\hbar} \Sigma^{\rho}_{\theta} } & ~=~
& e^{-i
\phi (\mu \rho - \frac{C_{\theta}}{\hbar} )} e^{\phi \partial_{\theta}}
\ , \nonumber \\
V^{\rho}(a) ~ := ~ e^{i \frac{a}{\hbar} \Sigma^{\rho}_{y} } & ~=~ & e^{i
a (\mu \theta + \frac{C_y}{\hbar}) } e^{a \partial_y } \ ,
\end{eqnarray}
so that
\begin{eqnarray}
\left\{ U^{\rho}(\phi) \psi \right\}(\theta, y) & = & e^{-i \phi (\mu
\rho - \frac{C_{\theta}}{\hbar})} \psi( \theta + \phi, y) \ , \nonumber \\
\left\{ V^{\rho}(a) \psi \right\}(\theta, y) & = &  e^{i a (\mu
\theta + \frac{C_y}{\hbar}) } \psi( \theta, y + a) \ . 
\end{eqnarray}

It is easily verified that
\begin{eqnarray}
U^{\rho}(\phi) V^{\rho}(a) & = & e^{ i a \mu \phi} ~ V^{\rho}(a) 
U^{\rho}(\phi) \ , \label{ProjAction}
\end{eqnarray}
and
\begin{eqnarray}
U^{\rho}(\phi) H^{\rho} & = & H^{\rho} U^{\rho}(\phi) \ , \nonumber \\
V^{\rho}(a) H^{\rho} & = & H^{\rho} V^{\rho}(a) \ . \label{Commut}
\end{eqnarray}

The equation (\ref{ProjAction}) above shows that the symmetry group $SO(2) 
\times \mathbb{Z}$ is projectively realized, independent of the choice of 
the constants $C_{\theta}, C_y$ and the choice of $\rho$ (compare
eq.(\ref{UCz})). 
The equations 
(\ref{Commut}) are {\em formally} valid even {\it without} imposition of
the condition $a \mu \in \mathbb{Z}$. As stated already in subsection 
\ref{Repn}, the Hamiltonian cannot distinguish between the cylinder and
its covering plane and so is insensitive to the discretizing condition.
The requirement that $V(a)$ respect the quasi-periodicity of the wave 
functions restricts $a$.

Now the fact that the translations along the $\theta$ directions form
the group $SO(2)$, requires that $U^{\rho}(2\pi)$ is the identity 
transformation. This together with the quasi-periodicity condition implies that 
$C_{\theta} = \hbar( \mu \rho - q) $ which fixes the definition of 
$U^{\rho}(\phi)$ appropriate for quasi-periodicity parameter $q$. 
There is no such requirement for $V^{\rho}(a)$, so $C_y$ is arbitrary and 
we may take it to be zero for notational convenience. 

The action of the group on the states of eq. (\ref{GroundState}) is obtained as:
\begin{eqnarray}
U^{\rho}(\phi) \Omega_{\rho, n}(\theta, y) ~~  & = &  ~~
e^{-i q \phi} \Omega_{\rho, n} (\theta + \phi, y) ~~~~~~~~~\, 
= ~~ e^{i n \phi} \Omega_{\rho, n}(\theta, y) \ , \nonumber \\
V^{\rho}(k \mu^{-1}) \Omega_{\rho, n}(\theta, y) ~~ & = & ~~
e^{i k \theta} \Omega_{\rho, n} (\theta, y + k \mu^{-1})
~~  =  ~~ \Omega_{\rho, n + k} (\theta, y ) \ .
\end{eqnarray}

The functions $\Omega_{\rho, n}(\theta, y)$ are thus eigenfunctions of 
action of the $SO(2)$ subgroup of the symmetry group while the action of
the $\mathbb{Z}$ subgroup mixes these eigenfunctions. 

One can also construct an eigenbasis such that the $\mathbb{Z}$ subgroup
acts diagonally and $SO(2)$ action mixes the basis states. These are
essentially obtained by a Fourier transform and are defined as,
\begin{eqnarray}
\tilde{\Omega}_{\rho, \xi} & ~ := ~ & \sum_{n \in \mathbb{Z}} e^{i n \xi}
\Omega_{\rho, n} ~~~~~ \forall ~ \xi \in [0, 2 \pi) ~;~ \nonumber \\
V^{\rho}(k \mu^{-1}) \tilde{\Omega}_{\rho, \xi} & ~ = ~ &
~ e^{- i k \xi} \, \tilde{\Omega}_{\rho, \xi} \nonumber \\
U^{\rho}(\phi) \tilde{\Omega}_{\rho, \xi} & ~ = ~ & ~ \tilde{\Omega}_{\rho, \xi + \phi} 
\end{eqnarray}

We conclude that, as in the case of the plane, the degeneracies are also
classified by the symmetry group, $\widetilde{SO(2) \times \mathbb{Z}}$.
We reproduce the same results as seen in the symmetry based approach of
the previous section.

In the present approach, we have two arbitrary parameters: $\rho$, labeling 
the gauge equivalence classes and $q$, the quasi-periodicity parameter.
Only one of them is physically relevant as can be seen from the following 
argument.

While the energy eigenvalues are independent of $\rho$ and $q$, the 
wave functions are not. In particular, for example,
$|\Omega_{\rho, 0}|^2(\theta, y)$ is a Gaussian in $y$, centred at $y = \rho
- q \mu^{-1}$. So $\rho$ and $q$ together, in this particular
combination, determine the peak of the Gaussian along the $y$-axis.
Since the choice of the origin along the $y$-axis is arbitrary, we can 
{\em always choose} it to coincide with this peak. Equivalently, given a 
quasi-periodicity parameter $q$, we can always choose the gauge equivalence
class labeled by $\rho = q \mu^{-1}$ {\em without any loss of generality}. 
Having made this choice, we are left with only the quasi-periodicity parameter
$q$ exactly as in the previous section. Notice that $\rho \to \rho + \mu^{-1}$
corresponds precisely to $q \to q + 1$ and $q$ is defined modulo integers. 
The $\rho$ is thus seen as parameterizing different, unitarily equivalent ways
of writing the wavefunction UR corresponding to different choices of origin
along the $y$-axis. 

Incidentally, by identifying $R \theta$ with $x$ and relaxing the identification
of $\theta = 0$ and $\theta = 2 \pi$, reproduces the planar case as
expected.

\section{Summary and Discussion}

We summarize our results and conclusions in the form of a series of
remarks.

(1) In the symmetry approach, reduction of the continuous translational
symmetry group to a discrete subgroup is a result of the non-existence of PURs
for the full symmetry group of the cylinder. In the Hamiltonian approach,
the same result is seen to be a consequence of the structure of the
group of gauge transformations and the corresponding existence of gauge 
inequivalent potentials. In both approaches, the basic cause of the
phenomenon is the non-contractible nature of the configuration space. 

(2) It is interesting to note that as $\hbar \to 0$ and/or $R
\to \infty$, the basic translation step, $ \mu^{-1} = 
\frac{\hbar}{e B R}$,  goes to zero implying essentially continuous 
translations being restored. The large $R$ limit takes us towards the
planar approximation while the vanishing $\hbar$ limit takes us to the
classical picture where continuous y-translations symmetry should be and
is recovered. This is understandable in view of the remark (3) in 
\ref{SymmDef}. The step size is relevant only for non-zero magnetic
field and the `free particle' limit is to be understood ab initio as
discussed in the remark (1) in \ref{SymmDef}. 

(3) It would be very interesting to experimentally verify the
discretization of the translation symmetry implied by quantization. Restoring 
the conventional units ($c \sim 3 \times 10^{10}$ cm/sec), the basic 
translation step selected by the system is $a = \mu^{-1} = \frac{\hbar c}{e B R}
~ \sim 6.6 \times 10^{-8} ~ B^{-1} R^{-1}$ centimeters. This is extremely
small eg. for $R \sim 1$ cm and $B \sim 1 $ gauss, the basic step is only of 
the order of an angstrom. Needless to say that some very clever
experimental tricks would be needed to verify the discretization.

(4) Similar conclusions hold in the case of the torus. The torus group
$SO(2) \times SO(2)$ has no non-trivial central extensions. Its
subgroups of the form $\mathbb{Z}_m \times \mathbb{Z}_n$ where $m, n$
are positive integers {\em do} have central extensions and they are
represented unitarily on the state space \cite{div-raj}. Intuitively one
expects the torus problem to effectively model a finite sample system.
Indeed in a recent paper \cite{Masson} discussing the finite geometry 
problem relevant for the quantum Hall effect, this expectation has been
corroborated to an extent. One may likewise expect our cylinder results
to model the problem on a strip.

(5) To summarize: At the level of a physical prediction, we see that for
the cylinder Landau electron, there is a {\it spontaneous} ``crystallization"
with lattice spacing along the axial direction being determined by the radius 
and the magnetic field. At the theoretical level, an important lesson is that
symmetries indicated by the classical magnetic field are not necessarily the 
ones realized quantum mechanically. One needs to pay attention to a complete 
gauge invariant specification of the system and then discover the allowed 
symmetries. The symmetry based approach, which is manifestly gauge
invariant, automatically gets the realizable symmetries directly by using only 
the general principles of symmetry realization in a quantum framework.

\begin{acknowledgments}
PPD expresses his appreciation to the Tata Institute of Fundamental
Research, Mumbai for hospitality. Both of us acknowledge discussions with
Prof. M. Barma and Prof. G. Baskaran.
\end{acknowledgments}

\end{document}